\documentclass[conference,letterpaper]{IEEEtran}

\usepackage[utf8]{inputenc} 
\usepackage[T1]{fontenc}
\usepackage{url}
\usepackage{ifthen}
\usepackage{cite}
\usepackage[cmex10]{amsmath} % Use the [cmex10] option to ensure complicance
\usepackage{amssymb}
\usepackage{float}
\usepackage{graphicx}
\usepackage{caption}
\usepackage{comment}
\usepackage{xcolor}
\usepackage{subcaption}

\begin{document}
\title{When and Where Faults Matter: A Study of Transient Errors in CKKS Multiplication}

\author{
    \IEEEauthorblockN{Vattana Chan\textsuperscript{3}, Matías Mazzanti\textsuperscript{1}, Karthik Swaminathan\textsuperscript{2}, Augusto Vega\textsuperscript{2}, Esteban Mocskos\textsuperscript{1}, Radha Venkatagiri\textsuperscript{3}\\}
    \IEEEauthorblockA{
        \textit{\textsuperscript{1}University of Buenos Aires, \textsuperscript{2}IBM T. J. Watson Research Center, \textsuperscript{3} Georgetown University}}
}

\maketitle
\begin{abstract}
Homomorphic Encryption (HE) is a privacy-preserving encryption paradigm that enables computation directly on encrypted data without requiring decryption. In this paper, we study errors in fully homomorphic encryption (FHE) computations, with a particular focus on server-side homomorphic multiplication in the unoptimized CKKS (Cheon--Kim--Kim--Song) scheme. We show that both the timing and the location of errors in the ciphertext components \(c_0\) and \(c_1\) have a significant impact on the correctness of the final FHE output.

% HE has been widely adopted in domains such as healthcare, finance, and government, where data confidentiality is critical. 
% The strong security guarantees of HE stem from the deliberate introduction of noise into ciphertexts, rendering otherwise simple mathematical
% problems computationally intractable, even for quantum adversaries.

% In practical deployments, especially in untrusted or
% resource-constrained server environments, hardware- and
% software-induced faults during homomorphic operations can
% interact with HE in unexpected ways. In some cases, such faults
% can evade traditional error detection and correction mechanisms,
% resulting in Silent Data Corruption (SDC) without triggering
% failures.

\end{abstract}

%\section{Introduction}
%- data independent, clear pattern for different computations in server
%- addition and rotation have different pattern than mult - how is it different?
%- mult dominates

\section{Introduction and Background}
\label{sec:introduction}
In the era of cloud computing \cite{GoogleTrace}, \cite{WarehouseScaleMachines}, security and privacy have become the primary
concerns, as users must entrust sensitive data to potentially untrusted servers. 
Homomorphic Encryption (HE) \cite{Gentry2009} is a cryptographic technique designed to address these concerns by enabling computation directly on encrypted data without revealing the underlying plaintext.

While the growing adoption of HE has driven substantial research into the design of new schemes \cite{Cheon2017, Halevi2019, BGN, Brakerski2014}, parameter optimization \cite{Tian2025LPHENN}, and
performance improvements \cite{F1, Cheetah, Poseidon, HEAP, TREBUCHET}, relatively little attention has been paid to the resilience of HE systems. CKKS \cite{Cheon2017, Cheon2019}; in particular, is a Fully Homomorphic Encryption (FHE) scheme that has received considerable attention from the research community due to its support for approximate arithmetic over real and complex numbers, making it especially well suited for privacy-preserving AI and machine learning workloads.

\begin{figure}[!hb]
  \centering
  \includegraphics[width=0.95\linewidth]{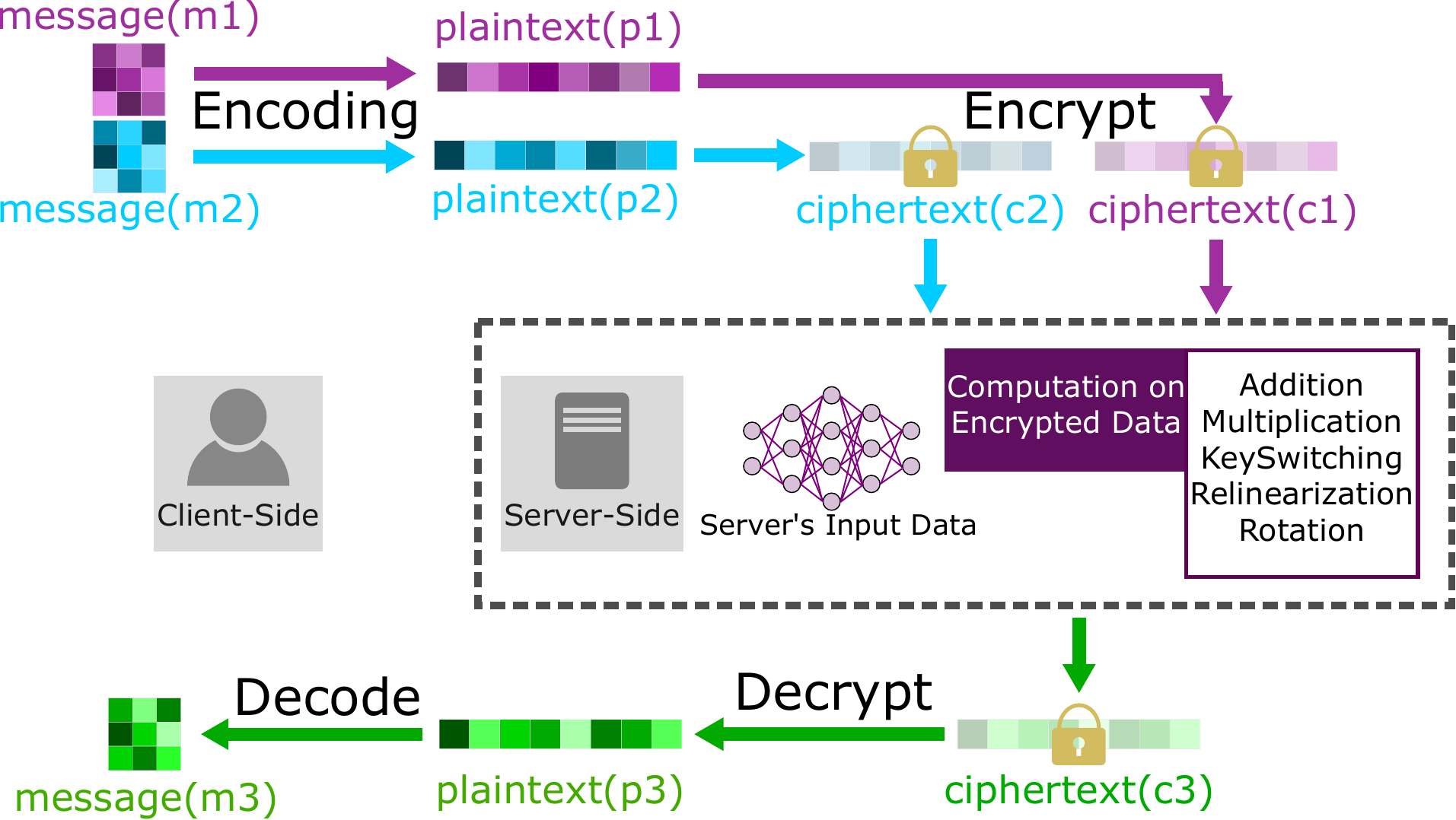}
  \caption[Typical FHE use case]{Overview of the end-to-end CKKS computational pipeline. Client-side encoding and encryption produce ciphertexts that are homomorphically evaluated on the server (where the errors studied in this work occur) before client-side decryption and decoding.\footnotemark}
  \label{fig:methodology}
\end{figure}
% \footnotetext{Adapted from~\url{https://openmined.org/blog/from-fully-homomorphic-encryption-to-silicon/}}

In CKKS, message is encoded using inverse Fast Fourier Transform (iFFT) and scaled by a scaling factor, transforming it into a plaintext (\textit{p}). During encryption, the plaintext then becomes a ciphertext. A ciphertext is represented as a pair of polynomials:
\begin{equation}\label{eq:ciph-mult}
  \begin{split}
      c &= (c_0, c_1)
  \end{split}
\end{equation}
where $c_0$ contains the encrypted message and $c_1$ is a public sampled key-dependent polynomial introduced during encryption. Decryption follows the standard RLWE \cite{Albrecht2015} paradigm: the plaintext is recovered by computing $c_0 + c_1 \times sk$, where $sk$ is the secret key. Through this operation, the randomness sampled during encryption is canceled, yielding the message together with the approximation error intrinsic to CKKS. Finally, decoding applies the Fast Fourier Transform (FFT) and rescales the decrypted plaintext to recover an approximation of the original input data.

\subsection{FHE Multiplication}

\label{subsection:fhe-mult}

 Ciphertext–ciphertext multiplication produces a three-term ciphertext:

  \begin{equation}\label{eq:ciph-mult}
  \begin{split}
      c^{mult}&=(c^{(1)}_0 c^{(2)}_0,\;c^{(1)}_0 c^{(2)}_1 + c^{(1)}_1 c^{(2)}_0,\; c^{(1)}_1 c^{(2)}_1) \\
      &= (d_0,\; d_1,\;d_2)
        \end{split}
  \end{equation}

CKKS then applies relinearization to convert this three-term ciphertext back into the standard two-term form. 
Using the evaluation key (evk), the third component is folded back as:
        \begin{equation}\label{eq:ciph-mult-relin}
            c^{mult}_{relin}\approx(d_0, d_1)+ d_2*evk\approx (c_0^{(3)},c_1^{(3)})
        \end{equation} 

    % \item Rotations (slot permutations) apply a cyclic shift to the encrypted packed
    % plaintext slots
    %        \[
    % c^{rot} = \text{Rot}_k(c) = KeySwitch(c,\text{ rk}_k) ,
    % \]
    % where $k$ is the rotation amount and $\text{rk}_k$ the corresponding
    % rotation key.
% \end{itemize}

In this work, we present an in-depth analysis of the resilience of homomorphic multiplication in CKKS in the presence of hardware-induced transient single-bit flip errors.
We show that:
1). homomorphic multiplication is highly susceptible to faults,
2). errors affecting ciphertext polynomials $c_0$ and $c_1$
follow distinct fault-propagation paths, and
3). the timing at which $c_0$ and $c_1$ encounter faults significantly influences the final decrypted output. These observations are supported by both theoretical analysis and empirical evaluation.

\section{Error Injection Methodology}
% \footnote{The FHE library used in this work (OpenFHE) detects data corruption and
% terminates execution with an assertion error whereas C-CKKS does not.}
Transient errors may occur during CKKS operations. These errors can lead to three distinctive outcomes: (i) the homomorphic
operation fails and the error is \textit{detected}, (ii) the operation completes successfully and the output is \textit{masked}, (iii) the operation completes successfully, but the error propagated and corrupts the final output, which is called \textit{Silent Data Corruption (SDC)} \cite{dixit2021silent}.

% Using our own CKKS implementation (C-CKKS) \footnote{C-CKKS is implemented based on OpenFHE \cite{AlBadawi2022}, HEaaN \cite{Cheon2017}, SEAL \cite{MicrosoftSeal} and PyFHE \cite{pyfhe}.} in C++, combined with the open-source error-injection tool LLTFI \cite{LLTFI}, we performed a systematic study of error propagation in CKKS Homomorphic Multiplication by varying critical FHE parameters such as modulus \textit{Q}, scaling factor \textit{$\Delta$}, ring dimension \textit{N}, including slots/gaps, generating us interesting insights on how FHE behaves under error-driven environment. After injecting each bit error, we run the full homomorphic encryption pipeline and evaluate the decoded output from the final stage by comparing it to the original data using the Maximum Relative Error Percentage (MREP).

Using our own CKKS implementation (C-CKKS) \footnote{C-CKKS is implemented based on OpenFHE \cite{AlBadawi2022}, HEaaN \cite{Cheon2017}, SEAL \cite{MicrosoftSeal} and PyFHE \cite{pyfhe}.} in C++ that provides us visibility into error propagation pathways, combined with error injection tool LLTFI \cite{LLTFI}, we injected single bit transient errors in different ciphertext polynomials during ciphertext-ciphertext multiplication as shown in Eq.~\ref{eq:ciph-mult} and Eq.~\ref{eq:ciph-mult-relin}. After injecting each bit error, we run the full homomorphic encryption pipeline and evaluate the decoded output from the final stage by comparing it to the original data using the Maximum Relative Error Percentage (MREP).

% Detecting such errors in HE applications is particularly challenging due to the noise-centric nature of HE schemes. 
% Ciphertexts are inherently noisy by construction, and additional hardware- or software-induced faults can be masked by this intrinsic noise, making detection extremely difficult. 

% For brevity, we limit our analysis to Vanilla homomorphic multiplication in this work and leave other operations for future, more detailed study. Previous studies \cite{mazzanti2025characterizingsensitivityindividualbit, vega2025quantifying} have suggested that homomorphic multiplication in RNS, NTT and RNS+NTT domains express very little to non-existent resilience at all.

To build a first-order understanding, we will only focus on homomorphic multiplication without optimizations such as NTT and RNS in this work, and leave other operations for future work.

% \begin{figure}[!ht]
%   \centering
%     \includegraphics[width=0.9\columnwidth]{figures/he_error_example.pdf}
% 	\caption{Illustrative scenario of a SDC case induced by a faulty CPU. The corrupted result incorporates both the HE error and the faulty hardware error.}
% 	\label{fig:he_error_example}
% \end{figure}

\section{Error Resilience Analysis}
\label{sec:error_resilience_analysis}
This section introduces preliminary results on CKKS resilience under transient errors during homomorphic multiplication.

% \textbf{FHE pipeline configurations.} We experimented FHE under various combinations of basic FHE computational operations such as No Computation (NoComp), Addition (Add), Multiplication (Mult), and Rotation (Rot).
% \begin{figure}[!ht]
%   \centering
%     \includegraphics[width=1\columnwidth]{figures/methodology.pdf}
% 	\caption{Error injection methodology.}
% 	\label{fig:methodology}
% \end{figure}

% \subsection{Inside Addition/Rotation}
% We observe that bit-flips below scaling factor (\textit{$\Delta$}) and above modulus (\textit{Q}) resilience lead to masked results for both $c_0$ and $c_1$, whereas bit-flips occur on gap and $\frac{N}{2}^{th}$ coefficients will always lead to fully masked results only in $c_0$, but not $c_1$. This is due to the fact that gaps as part of a encoded message are encapsulated into $c_0$. On the other hand, $c_1$ is a sampled-key; therefore, does not inherit the gaps and $\frac{N}{2}^{th}$ coefficients.

% The resilience profile of ciphertext within Homomophic Rotation mostly follows that of inside addition, even the ciphertexts are in key-switch step. When bit flips occur on $c_0$, we observe that gaps and $\frac{N}{2}^{th}$ coefficients are truncated (See Fig~ref/{}).

\subsection{Resiliency Analysis of FHE Multiplication}
\begin{figure}[!htbp]
  \centering
  \begin{minipage}[b]{\columnwidth}
      
    \includegraphics[width=\columnwidth]{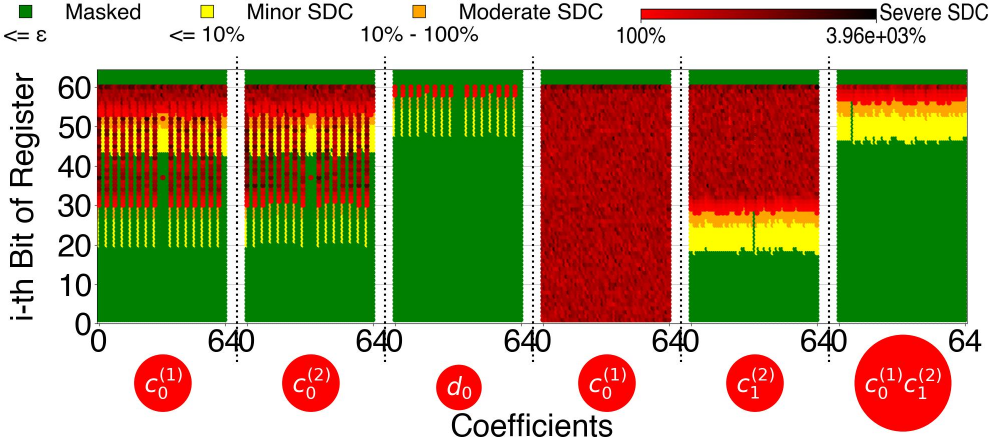}
    % \caption{Fault tree of $d_0$ and $d_1$ calculated with the same corrupted $c^{(1)}_0$}
    % \subcaption{}
    \label{fig:first-half-mult}
    
  \end{minipage}
  \hfill
  \begin{minipage}[b]{\columnwidth}
    \includegraphics[width=\columnwidth]{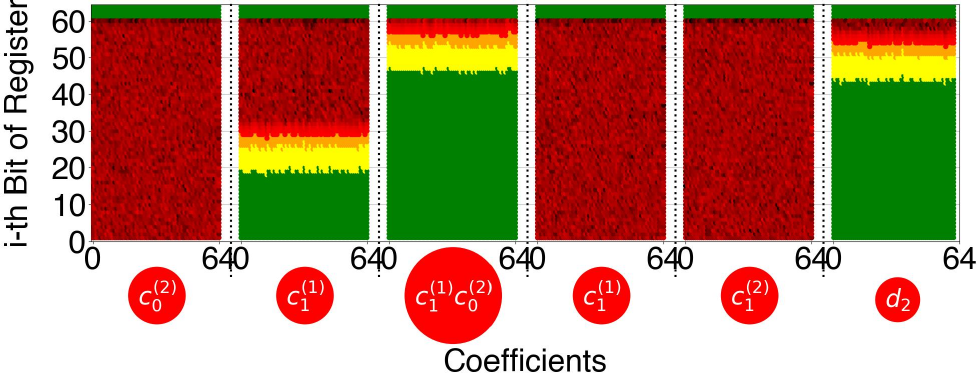}
    % \caption{Fault tree of $d_0$ and $d_1$ calculated with different versions of $c^{(1)}_0$. $d_0$ was calculated using error-free $c^{(1)}_0$ and $d_1$ was calculated using corrupted $c^{(1)}_0$.}
    % \subcaption{}
     \label{fig:second-half-mult}
  \end{minipage}

  \caption{Error characterization of bit-flips occurring on $c_0$ and $c_1$ inside Homomorphic Multiplication (Refer to \ref{subsection:fhe-mult}) with Gaps = 2, $\log\text{Q}$ = 60, $\log\Delta$ = 25, N = 64, $\varepsilon = 0.1$.}
  \label{fig:error-characterization}
\end{figure}

Due to the complexity of homomorphic multiplication, the timing and the location of bit-level faults can lead to remarkably different outcomes. In several cases, bit-flips in ciphertext components result in unexpected behaviors, such as the absence of the scaling factor~$\Delta$, the lack of a gap, or no $\frac{N}{2}^{th}$ coefficient resilience (See Fig~\ref{fig:error-characterization}) \footnote{The resilience of the scaling factor $\Delta$, the modulus $Q$, and the gap/slot structure, including the resilience profiles of state-of-the-art optimizations, is analyzed in detail in \cite{mazzanti2025characterizingsensitivityindividualbit, vega2025quantifying}.}. To understand these effects, we compared polynomial execution traces from error-free and fault-injected runs and used these comparisons to construct error propagation trees.

 \subsection{Error Propagation}
As shown in Equation~\ref{eq:ciph-mult} and \ref{eq:ciph-mult-relin}, both $c_0$ and $c_1$ from each ciphertext are each used twice in the computation of the final result: $c_0$ contributes to the formation of $d_0$ and $d_1$, while $c_1$ contributes to $d_1$ and $d_2$.
This inherent reuse induces a symmetric algebraic structure across the partial products, as detailed in Section~\ref{sec:symmetric-structure}.

We observe that when a bit-flip affects only one of these uses, this symmetry is broken, and the resulting polynomial no longer satisfies the cancellation properties required during decryption and decoding, leading to a silent data corruption (SDC). In contrast, when the same bit-flip is consistently injected into both uses of $c_0$ (or, analogously, both uses of $c_1$), the algebraic symmetry is preserved and the masking patterns previously described remain intact.

\subsection{Theoretical Explanation}\label{sec:symmetric-structure}

In order to gain a better understanding of what happens during ciphertext multiplication, and why resiliency is achieved in some cases but not in others, we analytically model the injected faults.

We model a fault as an error polynomial $\boldsymbol{e}_{i,j} \in \mathbb{Z}[X]$ whose only non-zero coefficient is the $i$-th one, corresponding to a bit flip at position $j$ of coefficient $i$ (using 0-based indexing and $i \in [0, N-1]$):
\begin{equation}
\boldsymbol{e}_{i,j} = 2^j X^{i}.
\end{equation}

The sign of this error depends on whether the bit is flipped from 0 to 1 or from 1 to 0; however, this distinction does not affect the following analysis.

In particular, we highlight the difference between corrupting $c^{(1)}_0$ in both $d_0$ and $d_1$, and corrupting it only in $d_1$.  
The case of $c^{(2)}_0$ is fully analogous.

We inject a fault in $c_0$, i.e., $c_0 \rightarrow c_0 + \boldsymbol{e}_{i,j}$, and expand the multiplication in Eq.~\ref{eq:ciph-mult-relin}:

\begin{equation}
    \begin{split}
      \textcolor{red}{\boldsymbol{c}_{\text{mult}}'} 
      =  (&(c^{(1)}_0 + \textcolor{red}{\boldsymbol{e}_{i,j}}) \times c^{(2)}_0 + d_0^{\text{evk}}, \\
       &c^{(1)}_1 \times c^{(2)}_0 + c^{(2)}_1 \times (c^{(1)}_0 + \textcolor{red}{\boldsymbol{e}_{i,j}}) + d_1^{\text{evk}}) \\
      & = \cdots \\
      & = (c^{(1)}_0{}^{m},\, c^{(1)}_1{}^{m}) 
        + \textcolor{red}{\boldsymbol{e}_{i,j} \times (c^{(2)}_0{}',\, c^{(2)}_1{}')} \\
      % & = \boldsymbol{c}_{\text{mult}} + \textcolor{red}{\boldsymbol{e}_{i,j} \times \boldsymbol{c}' }.
    \end{split}
\end{equation}

Thus, the resulting ciphertext corresponds to the correct multiplication outcome plus an additional error term given by the product of the injected fault and a valid ciphertext.

If we repeat the same analysis but inject the error only in $d_1$, the resulting expression is similar, yet it does not form a complete ciphertext. In this case, we obtain:
\begin{equation}
    \begin{split}
      \textcolor{red}{\boldsymbol{c}_{\text{mult}}'} 
      =  (&c_0 \times c_0' + d_0^{\text{evk}}, \\
      &c_1 \times c_0' + c_1' \times (c_0 + \textcolor{red}{\boldsymbol{e}_{i,j}}) + d_1^{\text{evk}}) \\
      & = \cdots = \boldsymbol{c}_{\text{mult}} + \textcolor{red}{\boldsymbol{e}_{i,j} \times (0,\, c_1')}.
    \end{split}
\end{equation}

%The key difference is that, in the first case, the injected fault results in a full ciphertext. 
%Consequently, during decryption \ref{sec:introduction}, the multiplication by the secret key induces the usual cancellation of terms, yielding an encoding of the now two messages plus the scheme-inherent approximation error.  

In the first case, the injected fault yields a valid ciphertext. During decryption (Section~\ref{sec:introduction}), the multiplication by the secret key induces the usual cancellation of terms, resulting in a decrypted value that can be interpreted as the intended product plus the scheme-inherent approximation error, and an additional low-magnitude term corresponding to the unaffected ciphertext.
While the explicit expansion involves many subterms ($d0, d1,d2,evk$, etc.), the overall effect on the decrypted value preserves the structure of the original multiplication. Consequently, the bit-flip fault pattern remains largely consistent with that previously characterized.

In contrast, when the fault affects only $d_1$, this cancellation does not occur, leaving residual randomness terms after decryption and therefore producing a significantly larger error.

% \section{Conclusion}
% In this study, we systematically examined the impact of hardware-induced faults on the reliability of homomorphic multiplication in CKKS on the server side.
% Our results show that homomorphic multiplication is highly susceptible to such faults.
% Moreover, errors affecting $c_0$ and $c_1$ follow distinct fault propagation patterns.
% Finally, the timing at which $c_0$ and $c_1$ are corrupted during multiplication plays a crucial role in determining the final computational outcome.

\bibliographystyle{IEEEtran}
\bibliography{refs}

\end{document}